\documentclass[letter]{aa}

\usepackage{graphicx}
\usepackage{txfonts}
\usepackage[colorlinks=true,
            urlcolor=blue,
            linkcolor=blue,
            citecolor=blue]
            {hyperref}
\usepackage{tabu}
\usepackage{ulem}
\usepackage{multirow} 
\usepackage{gensymb}
\newcommand{\Ha}{H$\alpha$}
\newcommand{\Hb}{H$\beta$}

\newcommand{\Lya}{Ly$\alpha$}

\newcommand{\kms}{km\,s$^{-1}$}

\begin{document}

   \title{
   First reverberation mapping of a Bowen fluorescence line
   }

      \author{Martin W. Ochmann \inst{1,\thanks{Corresponding author: \texttt{martin.ochmann@uni-goettingen.de}}},
           Edward~M.~Cackett \inst{2},
           Lukas~Diehl \inst{1},
           Keith~Horne \inst{3},
           Malte~A.~Probst \inst{1},
           Wolfram~Kollatschny \inst{1}
           }

   \institute{
          Institut f\"ur Astrophysik und Geophysik, Universit\"at G\"ottingen,
          Friedrich-Hund-Platz 1, 37077 G\"ottingen, Germany
          \and
          Department of Physics and Astronomy, Wayne State University, 666 W. Hancock Street, Detroit, MI 48201, USA
          \and
          SUPA Physics and Astronomy, University of St. Andrews, Fife, KY16 9SS, United Kingdom
          }

  \date{Received 07 October 2025 / Accepted 17 December 2025}

 \abstract{
 Reverberation mapping (RM) is a powerful tool to determine the extent, structure, and kinematics of the broad-line region (BLR) of active galactic nuclei (AGNs). So far, RM of the BLR has only been performed for recombination lines responding to the varying ionizing continuum.
}
{We tested whether \ion{O}{i}$\,\lambda8446$, attributed to Bowen fluorescence driven by Ly$\beta$ pumping, varied on short (day- to week-long) timescales during a 2016 Hubble Space Telescope/Space Telescope Imaging Spectrograph (HST/STIS) campaign of NGC\,4593, and examined how it relates to other emission lines and the ionizing UV continuum.
}
{We quantified the variability of \ion{O}{i}$\,\lambda8446$ by its root-mean-square (rms) amplitude. We then extracted integrated light curves of \ion{O}{i}$\,\lambda8446$ and other UV and optical emission lines, and compared them with each other and the UV continuum light curve using correlation analyses. In addition, we used archival near-infrared spectra to assess the dominant excitation mechanism of \ion{O}{i}$\,\lambda8446$.
}
{We detect, for the first time, variability in \ion{O}{i}$\,\lambda8446$  on day timescales. The fractional rms amplitude is $\sim 4$\% over the 4-week campaign. The \ion{O}{i}$\,\lambda8446$ light curve reverberates with a delay of $\sim 2.5$ days relative to \Lya{}, used as a proxy for Ly$\beta$,  detected at a false-alarm probability of 0.6\% (significance of $\sim 2.8\sigma$) under our adopted null hypothesis. It closely tracks \Ha{} with only a minor additional delay of $\sim0.3$\,days, placing its emission region at essentially the same distance as the Balmer-line weighted BLR. Line ratios indicate that Ly$\beta$ pumping is the dominant excitation mechanism for \ion{O}{i}$\,\lambda8446$.
}
{Our results establish \ion{O}{i}$\,\lambda8446$  as the first Bowen fluorescence line to be reverberation-mapped, responding directly to variations in the Ly$\beta$ flux. We propose that in future campaigns targeting AGNs with larger BLRs, \ion{O}{i} could enable dual-driver RM using both the continuum and the pumping line as drivers.
}


  \titlerunning{First reverberation mapping of a Bowen fluorescence line}
  \authorrunning{M.W. Ochmann et al.}

  \maketitle
  \nolinenumbers
%
%

%**********************************************************************************
%
\section{Introduction}\label{sec:introduction}
%
%**********************************************************************************
Variability across all wavelength bands is a characteristic feature of active galactic nuclei (AGNs) and was already recognized in the earliest studies of these objects \citep{matthews63, fitch67}. It occurs in all spectral regions on timescales ranging from hours to weeks or even years \citep[][]{ulrich97} and has proven highly valuable for probing the inner structure of AGNs. In particular, the broad-line region (BLR) is traditionally studied through reverberation mapping (RM), which uses the delayed response of recombination emission lines to continuum variations to investigate the structure and kinematics of the emitting gas \citep{blandford82, peterson93, cackett21}. Usually, the bluest available continuum is used as a proxy for the ionizing continuum, which is assumed to drive the line variability.

Balmer lines have been extensively monitored in the past due to their accessibility in optical spectra for nearby AGNs, and have been used with great success to determine the size \citep[][]{peterson91, kaspi00, grier17} -- and even the kinematics \citep[][]{ulrich96,kollatschny03,bentz09,horne21} -- of the BLR\footnote{We note that RM has been applied not only to the BLR, but also to the accretion disk \citep[e.g.,][]{fausnaugh16,cackett18}, the dusty torus \citep[e.g.,][]{koshida14}, and the coronal line region \citep[e.g.,][]{oknyanskii91,yin25}.}. However, much less attention has overall been given to other lines, high-ionization lines such as \ion{C}{iv}$\,\lambda1548$ \citep[e.g.,][]{clavel91, goad16, grier19} or low-ionization lines such as \ion{Mg}{ii}$\,\lambda2800$ \citep[e.g.,][]{shen16,czerny19,prince23}, mainly due to observational limitations. Other low-ionization lines, such as \ion{O}{i}$\,\lambda8446$ or the \ion{Ca}{ii} triplet $\,\lambda\lambda8498,8542,8662$, are entirely absent from RM campaigns, despite their strong diagnostic power for the physical conditions and excitation mechanisms in the BLR \citep[e.g.,][]{grandi80, persson88, ferland89, joly89, rodriguez-ardila02b, matsuoka07, landt08, matsuoka08, marziani13, panda20, martinez-aldama21}.

Among these lines, the \ion{O}{i}$\,\lambda8446$ transition is of particular interest,  as it can be enhanced through Bowen fluorescence driven by Ly$\beta$ pumping \citep{bowen47,netzer76,kwan81}. Recently, Bowen fluorescence lines have gained more attention in AGN studies following the detection of their variability in spectra of flaring AGNs \citep{trakhtenbrot19a,makrygianni23,ochmann24,baldini25,sniegowska25}. However, dedicated short-term (days to weeks) monitoring campaigns targeting any Bowen fluorescence line are still absent. Here we present the first measurement of \ion{O}{i}$\,\lambda8446$ variability and lag in NGC\,4593 from a RM campaign with the Hubble Space Telescope/Space Telescope Imaging Spectrograph (HST/STIS), aiming to examine its relation to other emission lines and to the ionizing UV continuum.

%**********************************************************************************
%
\section{Observations}\label{sec:observations}
%
%**********************************************************************************

For this study we utilized 26 HST/STIS observations of NGC\,4593 obtained during 2016 July 12 and August 6 with nearly daily cadence. Analysis of the continuum reverberation lags obtained from the same HST data used here were presented in \citet{cackett18}, where full details of the instrumental setup and data reduction are given. Analysis of the {\it Swift} data from the same campaign was presented in \citet{mchardy18}. 

The original spectra presented in \citet{cackett18} are not corrected for small variations in the wavelength scale between observations. For the purpose of emission-line RM, we corrected these variations in each individual spectrum by performing minor wavelength shifts ($< 1$\,\AA{}) and scaling the [\ion{O}{iii}]$\,\lambda5007$ flux, which is assumed to be constant on timescales of years, to a common value of {$(106 \pm 5) \times 10^{-15}$ erg s$^{-1}$ cm$^{-2}$ , thereby minimizing the [\ion{O}{iii}]$\,\lambda5007$ residual in the root-mean-square (rms) spectrum. This procedure is standard in optical spectroscopic RM campaigns \citep[e.g.,][]{vangroningen92}. The adjustments were applied only to the G430L and G750L parts of the spectra, as intercalibrations based on narrow lines are only reliable over a limited spectral range. 

%**********************************************************************************
%
\section{Results}\label{sec:results}
%
%**********************************************************************************

%**********************************************************************************
%
\subsection{The variability of \ion{O}{i}$\,\lambda8446$ }\label{sec:res_variability_OI8446}
%
%**********************************************************************************

We calculated the mean and rms spectra from the 26 HST/STIS observations. Figure~\ref{fig:NGC4593_avg_rms} shows the resulting mean and rms spectra, together with the interpolated linear pseudo-continuum beneath \ion{O}{i}$\,\lambda8446$ used for further analysis. The rms spectrum is scaled and shifted for clarity. The most prominent lines in the spectra are indicated. The inset in the top right corner shows a zoom-in on the near-infrared \ion{O}{i} and \ion{Ca}{ii} triplet complex. The rms spectrum reveals variability in \ion{O}{i}$\,\lambda8446$ above the continuum level. To quantify this variability, we determined the fractional rms amplitude, $F_{\rm rms}/F_{\rm mean}$, of the integrated \ion{O}{i}$\,\lambda8446$ line flux within the wavelength range 8380–8498\,\AA{}. This interval includes most of the \ion{O}{i}$\,\lambda8446$ flux, while limiting the potential contribution from \ion{Ca}{ii}$\,\lambda\lambda8498,8542$ to $\lesssim20$\% \citep[see][]{ochmann25}. With this definition, we find a fractional rms amplitude of $\sim 4$\%. This represents the first robust detection of day-scale, stochastic variability in \ion{O}{i}$\,\lambda8446$ during a non-transient AGN monitoring campaign.

\begin{figure*}[h!]
    \centering
    \includegraphics[width=0.99\textwidth,angle=0]{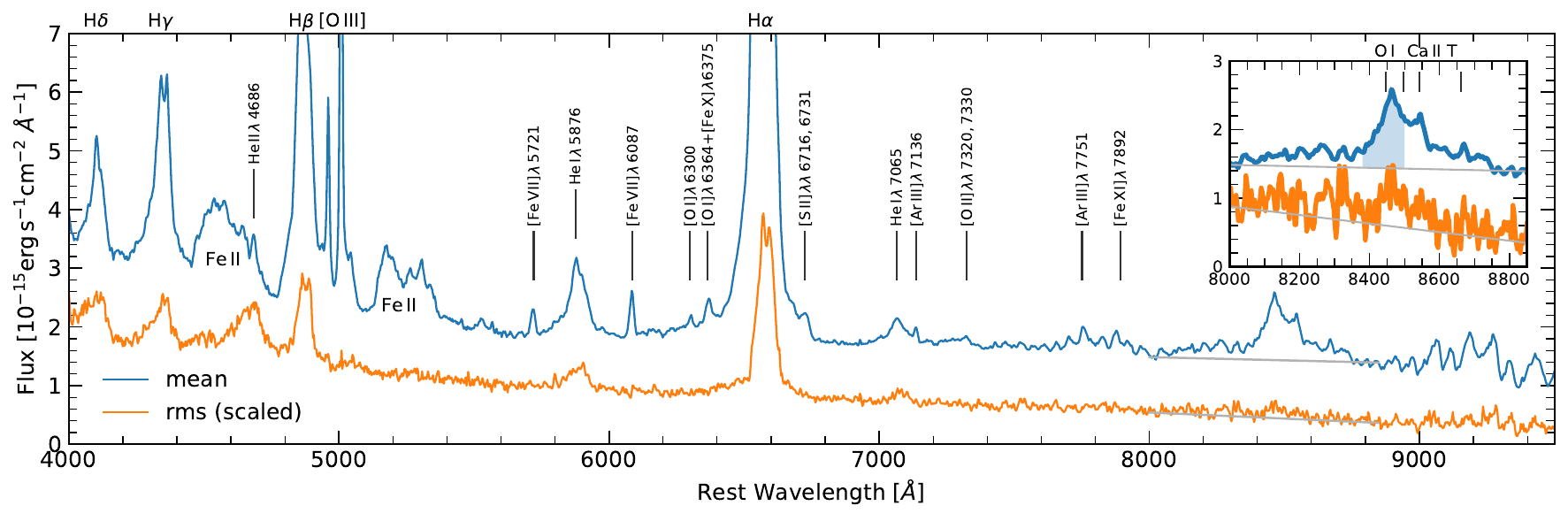}
    \caption{Optical mean (blue) and rms (orange) spectra from the HST/STIS campaign. The rms spectrum is scaled ($\times$5) and shifted in flux ($- 0.25 \times 10^{-15}$ erg s$^{-1}$ cm$^{-2}$ \AA{}$^{-1}$) to highlight weak line features. The linear pseudo-continuum beneath the \ion{O}{i}\,$\lambda8446$ and the \ion{Ca}{ii} triplet complex is shown for orientation. The inset shows this spectral region in greater detail, with the rms spectrum scaled ($\times$15) and shifted ($- 1.5 \times 10^{-15}$ erg s$^{-1}$ cm$^{-2}$ \AA{}$^{-1}$). The \ion{O}{i}$\,\lambda8446$ integration area is shaded in blue (see Sect.~\ref{sec:res_response_OI8446}).}
\label{fig:NGC4593_avg_rms}
\end{figure*}
%

%**********************************************************************************
%
\subsection{The response of \ion{O}{i}$\,\lambda8446$}\label{sec:res_response_OI8446}
%
%**********************************************************************************
To examine the expected connection between \ion{O}{i}$\,\lambda8446$ and Ly$\beta$ pumping, we cross-correlated the \ion{O}{i} light curve with those of \Lya{}, \Ha{}, and the \textit{Swift} UVW2 continuum \citep{mchardy18} using the interpolated cross-correlation function \citep[ICCF;][]{gaskell87}. Since Ly$\beta$ lies outside the spectral range of the HST/STIS data, we used \Lya{} as a proxy, as both lines are expected to arise under similar physical conditions. The normalized light curves as well as the resulting CCFs are shown in Fig.~\ref{fig:NGC4593_OI_CCFs}. The time delay between light curves was determined from the centroid, $\tau_{\rm cent}$, of the  upper 20\% of the CCF. The error margins were estimated using the flux randomization and random subset selection (FR and RSS) method with $10\,000$ independent realizations \citep{peterson98}.

\begin{figure}[h!]
    \centering
    \includegraphics[width=0.5\textwidth,angle=0]{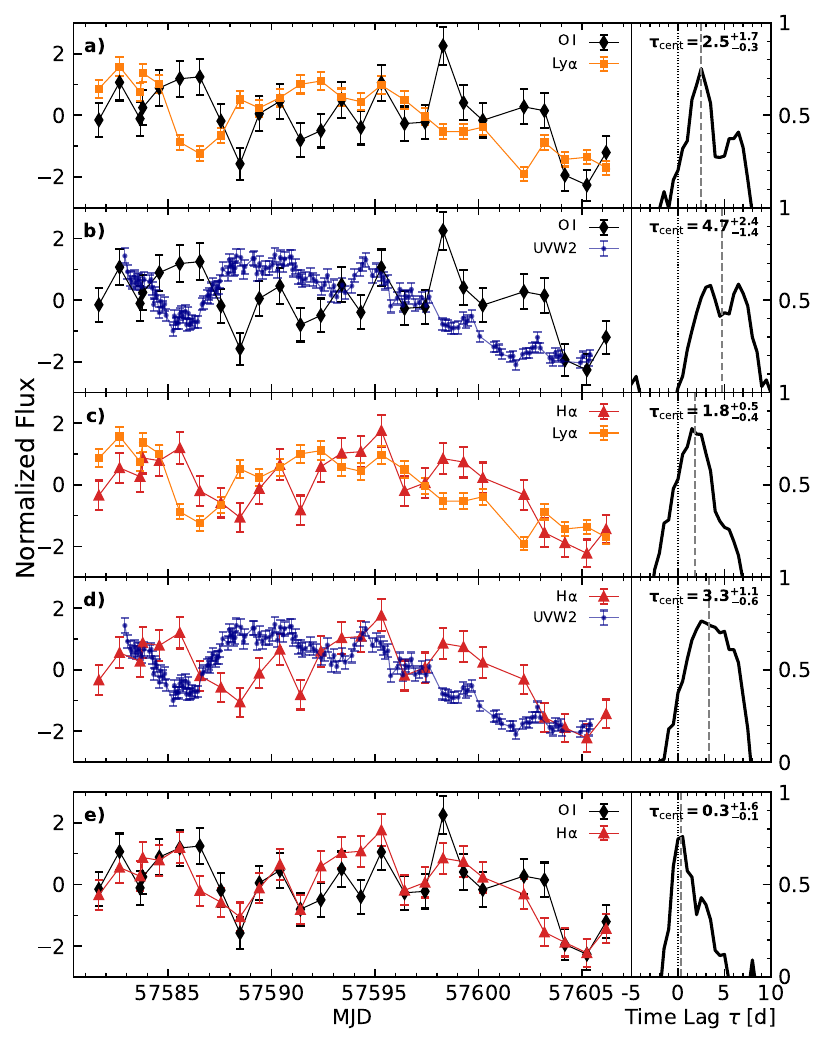}
    \caption{Comparison of normalized light curves (left panels) of the emission lines \ion{O}{i}\,$\lambda8446$, \Lya{}, and \Ha{}, together with the UVW2 continuum, with CCFs (right panels) of each pair, showing correlated variability and time delays between the individual emitting components. The CCFs are plotted between 0 and 1. The dashed line indicates $\tau_{\rm cent}$.}
\label{fig:NGC4593_OI_CCFs}
\end{figure}

Overall, we find that the line and continuum light curves are similar in shape but shifted relative to each other. All correlations are strong, with $r_{\rm max}\sim 0.75-0.80$, except for \ion{O}{i} versus UVW2, where we find a lower value of $r_{\rm max}\sim 0.60$. \ion{O}{i}$\,\lambda8446$ lags \Lya{} and the UVW2 continuum by $2.5^{+1.7}_{-0.3}$ and $4.7^{+2.4}_{-1.4}$\,days, respectively (Fig.~\ref{fig:NGC4593_OI_CCFs} a,b). In comparison, \Ha{} lags \Lya{} and the UVW2 continuum by $1.8^{+0.5}_{-0.4}$ and $3.3^{+1.1}_{-0.6}$\,days, respectively, corresponding to time lags shorter by about $0.5$–$1$\,days,  although the differences are not significant within the $1\sigma$ uncertainties  (Fig.~\ref{fig:NGC4593_OI_CCFs} c,d). When comparing the light curves of \ion{O}{i}$\,\lambda8446$ and \Ha{} directly, we find good agreement overall, with a tendency for \ion{O}{i} to vary later than \Ha{} by $\sim 0.3$\,days (Fig.~\ref{fig:NGC4593_OI_CCFs} e). To test the significance of the \ion{O}{i} lags, we performed a Monte-Carlo permutation test in which we randomly permuted the \ion{O}{i} fluxes among the observed epochs under the null hypothesis of no correlation (see Sect.~\ref{sec:app_significance_lag_detection} for details). This test yields a false-alarm probability of $\sim 0.6\%$ (significance of $\sim2.8\sigma$) for the \ion{O}{i}–\Lya{} pair, and a false-alarm probability of $\sim 11.6\%$ (significance of $\sim 1.6\sigma$) for the \ion{O}{i}–UVW2 pair, consistent with a substantially weaker and less robust correlation in the latter case.

%**********************************************************************************
%
\section{Discussion}\label{sec:discussion}
%
%**********************************************************************************

%**********************************************************************************
%
\subsection{The excitation mechanism and origin of variability}\label{sec:disc_excitation_mechanism}
%
%**********************************************************************************
Recombination, collisional excitation, continuum fluorescence, and Bowen fluorescence driven by Ly$\beta$ pumping are all possible excitation mechanisms of \ion{O}{i}$\,\lambda8446$ \citep[see][]{grandi80}. To identify the dominant mechanism in NGC\,4593, we examined the presence of other permitted \ion{O}{i} emission lines in the mean HST/STIS spectrum. We also analyzed near-infrared SpeX and MUSE spectra of NGC\,4593 from 2006 and 2019 presented by \citet{landt08} and \citet{ochmann25}, respectively.\footnote{We consider it justified to use spectra from epochs separated by several years for line diagnostics, as the spectra of NGC\,4593 over the past 20 years show only typical variability, with no evidence for fundamental changes in the continuum flux driving the broad-line emission \citep[see][]{denney06,landt08,barth15,cackett18,ochmann25}.}

We exclude recombination as a significant contributor to the production of \ion{O}{i} based on the absence of detectable quintet lines with comparable strength to that of \ion{O}{i}$\,\lambda8446$ \citep[see][]{grandi75}. In particular, recombination would imply a relative strength of $\lambda7774/\lambda8446 \approx 1.1 - 1.7$ \citep{grandi80,landt08}. We derive an upper limit of $\lambda7774/\lambda8446 = 0.2$ from the MUSE spectrum, and this value is only reached if the line previously identified as \ion{He}{i}$\,\lambda 7816$ is almost fully attributed to the right peak of a double-peaked \ion{O}{i}$\,\lambda7774$. While the ratio of $\lambda7774/\lambda8446 \leq 0.2$ could imply the presence of collisional excitation for which a ratio of $\approx 0.3$ is expected \citep{grandi80}, continuum fluorescence can be excluded by the complete absence of \ion{O}{i}$\,\lambda13165$ in the SpeX spectrum \citep[see][]{grandi75,grandi80}. To test the presence of Ly$\beta$ pumping, we calculated the photon ratio n$(\lambda11287)$/n$(\lambda8446)$ after decomposition of the \ion{O}{i} and \ion{Ca}{ii} triplet complex following the procedure described by \citet{ochmann25} and using the same \ion{Ca}{ii}$\,\lambda8662$ template. We find a photon ratio of $\sim 0.8$, close to unity as would be the case for Ly$\beta$ as the sole excitation mechanism \citep[e.g.,][]{rudy89}. We conclude that Ly$\beta$ is indeed the dominant excitation mechanism\footnote{We also calculated the photon ratio n$(\lambda1304)$/n$(\lambda8446)$ from the HST/STIS spectrum and obtain only $\sim 0.1$, well below the expected one-to-one relation. This discrepancy may arise from absorption features near \ion{O}{i}$\,\lambda1304$, possible reddening, and destruction mechanisms that can deplete \ion{O}{i}$\,\lambda1304$ photons \citep[see][]{kwan81,grandi83}, effectively breaking the expected relation. No reverberating signal in \ion{O}{i}$\,\lambda 1304$ is found.}, and the  discrepancy from unity can be explained by a secondary contribution from collisional excitation, similarly to what was found for other sources \citep{rodriguez-ardila02b,matsuoka07,landt08,tripodi25}. Based on our line diagnostics, we suggest that Ly$\beta$ pumping is the driver of the detected day-scale variability in \ion{O}{i}$\,\lambda8446$; that is, the variability of \ion{O}{i}$\,\lambda8446$ is caused by variability in the incident Ly$\beta$ flux.

%**********************************************************************************
%
\subsection{The emission region}\label{sec:disc_emission_region}
%
%**********************************************************************************
From the cross-correlation analysis in Sect.~\ref{sec:res_response_OI8446}, we obtain an H$\alpha$-weighted BLR radius of $3.3^{+1.1}_{-0.6}$\,light-days in NGC\,4593, based on the lag relative to UVW2. This value is in excellent agreement with BLR radii of this source determined in earlier studies \citep[][]{dietrich94,kollatschny97,denney06, williams18}. In comparison to \Ha{}, \ion{O}{i}$\,\lambda8446$ has a slightly higher delay of $4.7^{+2.4}_{-1.4}$\,days relative to the UVW2 continuum, consistent within the uncertainties.\footnote{We note that the correlation between \ion{O}{i} and UVW2 is weaker ($r_{\rm max} \sim 0.60$ vs. $\sim 0.80$ for \Ha{}), and the CCF peak is less well defined, providing additional indication that the continuum is not the true driver of \ion{O}{i} but that the correlation rather reflects the general similarity of all light curves.} At first glance, this agrees with studies indicating that \ion{O}{i}, together with \ion{Fe}{ii} and \ion{Ca}{ii}, is emitted at larger radii than the hydrogen lines \citep[e.g.,][]{persson88,rodriguez-ardila02a,marinello16}, based on the typically larger widths of the hydrogen lines in some objects, in particular in Pop. B sources \citep{martinez-aldama15,martinez-aldama21}. However, we argue that the delays relative to UVW2 should not be compared directly, since we identified Bowen fluorescence driven by Ly$\beta$ pumping as the primary excitation mechanism of \ion{O}{i} in NGC\,4593. Accordingly, the driving light curve of \ion{O}{i} is \Lya{} as a proxy for Ly$\beta$, and the distance to the ionizing source -- assuming a planar BLR geometry, as is supported by the double-peaked broad emission lines in NGC\,4593 \citep{ochmann25} -- corresponds to the delay of \Lya{} relative to UVW2 plus the delay of \ion{O}{i} relative to \Lya{}. For the delay of \Lya{} relative to UVW2, we obtain $1.0^{+0.4}_{-0.2}$\,days. The distance of the \ion{O}{i} region from the continuum source is therefore $\sim 3.5$\,days. This corresponds to the distance of the \Ha{} region from the continuum source as established in Sect.~\ref{sec:res_response_OI8446}. 

We find a very good correlation ($r_{\rm max}\sim0.80$}) between the \Ha{} and \ion{O}{i}$\,\lambda8446$ light curves (see Fig.~\ref{fig:NGC4593_OI_CCFs} e), with only a minor time shift of $\sim0.3$\,days. This supports the conclusion that, within the uncertainties, both lines in NGC\,4593 are emitted at essentially the same distance from the continuum source. A detailed study of emission-line profiles in this galaxy by \citet{ochmann25} revealed double-peaked structures in both \Hb{} and \ion{O}{i}$\,\lambda8446$, with \Hb{} broader by $\sim25$\%  compared to \ion{O}{i}$\,\lambda8446$ \citep[4500\,\kms{} vs. 3580\,\kms{}; see also][]{marinello16}. Notably, the red peak of \Hb{} is shifted toward lower velocities by $\sim200$\,\kms{}. If the line widths are governed primarily by virial motions, as is often assumed in profile comparison studies of \ion{O}{i}/\ion{Ca}{ii} and \Hb{}/\Ha{}, this broadening and peak shift cannot be explained by placing the Balmer phase of the BLR closer to the ionizing source within the same disk-like configuration, as this would naturally result in larger peak separations. We instead propose an explanation involving a vertically stratified, disk-like BLR with scale-height-dependent turbulence \citep{goad12, kollatschny13a, kollatschny13b}, in which the different broad emission lines arise in distinct layers parallel to the disk plane. A similar configuration -- a plane-parallel distribution of emitting gas above and below the disk -- was already discussed by \citet{martinez-aldama15} in the context of \ion{O}{i}, \ion{Ca}{ii}, and \Hb{}, and was  invoked by \citet{ochmann24} to explain the differing profile widths and shapes of these lines in NGC\,1566.

We note that the detection of Ly$\beta$/\Lya{} as variability drivers in AGN emission lines, together with the proposed vertically stratified BLR structure, could be exploited in future campaigns to perform dual-driver RM. For the small BLR size in NGC\,4593, differential light-travel time delays between the cascaded responses are too short to be resolved even with daily cadence. In other words, we cannot separate the delay chain from the continuum through \Lya{} to O\,I from that of the continuum to \Ha{}, or the relative delay between H$\alpha$ and O\,I. In sources with much larger BLRs, however, such delays may become accessible, provided that the different excitation channels operate on spatial scales comparable to the observational cadence.

%**********************************************************************************
%
\section{Conclusions}\label{sec:conclusions}
%
%**********************************************************************************
In this study, we present the first RM analysis of a Bowen fluorescence line, \ion{O}{i}$\,\lambda8446$, based on an HST/STIS monitoring campaign of NGC\,4593. We quantified the variability of \ion{O}{i}$\,\lambda8446$, examined its relation to other emission lines and to the ionizing UV continuum, and determined the dominant excitation mechanism. Our main results are as follows:

\begin{enumerate}
    \item We detect, for the first time, short-term (day-scale) variability in \ion{O}{i}$\,\lambda8446$ during a non-transient AGN monitoring campaign, with a fractional rms amplitude of $\sim 4$\%. Line diagnostics show that Ly$\beta$ pumping is the dominant excitation mechanism.  

    \item The \ion{O}{i}$\,\lambda8446$ light curve reverberates with a delay of $\sim 2.5$ days relative to \Lya{}, used here as a proxy for Ly$\beta$, detected at a false-alarm probability of 0.6\% (significance of $\sim 2.8\sigma$) under our adopted null hypothesis. It closely tracks \Ha{} with only a minor additional delay ($\sim0.3$\,days), placing its emission region at essentially the same distance as the Balmer-line weighted BLR.

\end{enumerate}
Our results establish \ion{O}{i}$\,\lambda8446$  as the first Bowen fluorescence line to be reverberation-mapped, responding directly to variations in Ly$\beta$ flux. We propose that in future campaigns targeting AGNs with larger BLRs, \ion{O}{i} could enable dual-driver RM using both the continuum and the pumping line as drivers, offering a new way to probe the structure of the BLR.

\begin{acknowledgements}
The authors thank the anonymous referee for helpful comments that strengthened the paper. MWO acknowledges the support of the German Aerospace Center (DLR) within the framework of the ``Verbundforschung Astronomie und Astrophysik'' through grant 50OR2305 with funds from the BMWK. EMC gratefully acknowledges support for program number 14121 provided by NASA through a grant from the Space Telescope Science Institute, which is operated by the Association of Universities for Research in Astronomy, Incorporated, under NASA contract NAS5-26555. MAP and WK greatly acknowledge support by the DFG grants KO 857/35-1 and KO 857/35-2.

\end{acknowledgements}

\bibliographystyle{aa}
\bibliography{literature}

\appendix

\section{Significance of the \ion{O}{i} lag detection}
\label{sec:app_significance_lag_detection}

\begin{figure}[h!]
    \centering
    \includegraphics[width=0.5\textwidth,angle=0]{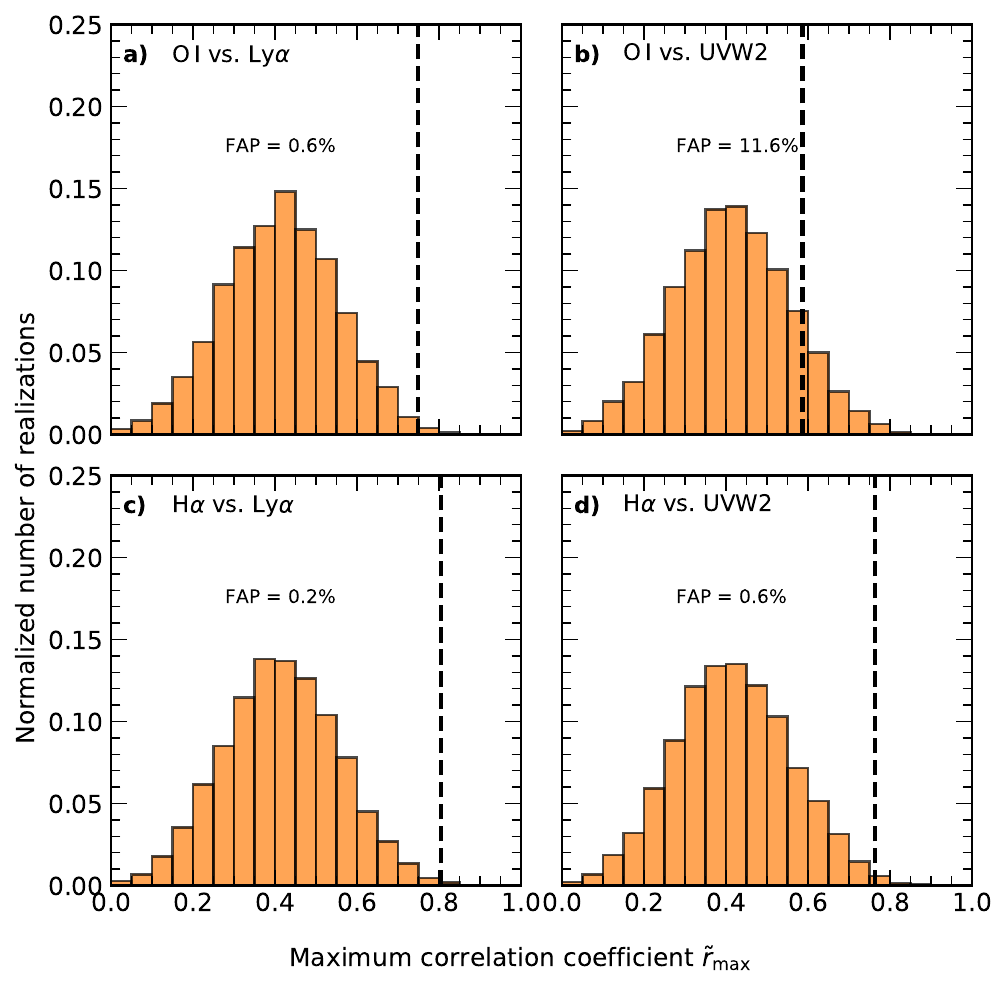}
    \caption{Distribution of $\tilde{r}_{\rm max}$ for 10\,000 realizations without physical correlation between the line light curve and the supposed driving light curve. The maximum correlation coefficient $r_{\rm max}$ obtained from the ICCF analysis (see Sect.~\ref{sec:res_response_OI8446}) is indicated as a dashed black line.}
\label{fig:NGC4593_FAP}
\end{figure}

To assess the significance of the \ion{O}{i}$\,\lambda8446$ lag detections with respect to \Lya{} and the UVW2 light curve, we calculated the false alarm probability (FAP), that is, the likelihood that the correlation coefficients $r_{\rm max}$ reported in Sect.~\ref{sec:res_response_OI8446} arise from random, uncorrelated light curves. To this end, we created 10\,000 surrogate light curves of \ion{O}{i}$\,\lambda8446$ and \Ha{} under the null hypothesis of no correlation. In each realization, we randomly permuted the 
\ion{O}{i}$\,\lambda8446$ and \Ha{} fluxes among the observed epochs and added Gaussian perturbations according to the measurement uncertainties, thereby preserving the flux distribution, sampling pattern, and uncertainty properties, while destroying any physical correlation with \Lya{} and UVW2, respectively. At the same time, we keep the \Lya{} and UVW2 light curves fixed, thereby
preserving the observed red-noise variability of the driving light curves. For each realization, we then recorded $\tilde{r}_{\rm max}$ of the ICCF, and the FAP was calculated from the number of realizations for which $\tilde{r}_{\rm max} > r_{\rm max}$. The distributions of $\tilde{r}_{\rm max}$ for each light-curve pair are shown in Fig.~\ref{fig:NGC4593_FAP}.

The surrogate $\tilde{r}_{\rm max}$ distributions for all four line–driver combinations are very similar in shape and are mainly determined by the sampling pattern and lag range. For the pairs
\ion{O}{i}--\Lya{}, \Ha{}--\Lya{}, and
\Ha{}--UVW2 we obtain FAP(\ion{O}{i}--\Lya{}) = 0.6\%,
FAP(\Ha{}--\Lya{}) = 0.2\%, and FAP(\Ha{}--{\rm UVW2}) = 0.6\%, indicating that in these three cases the correlations are detected at $>99\%$ confidence and are unlikely to arise from uncorrelated light curves. For the \ion{O}{i}$\,\lambda8446$–UVW2 pair we find FAP(\ion{O}{i}--UVW2) = 11.6\%, implying that the corresponding correlation is substantially less significant than for the other pairs and that the \ion{O}{i}-UVW2 lag should be regarded as comparatively weak and less robust.

We note that the null test preserves the observed flux distribution, sampling, and measurement uncertainties, but does not model the red-noise variability of the line light curves. Given the short baseline and modest variability, more sophisticated models (e.g., a damped random walk) are poorly constrained and thus not pursued. 

In addition to the calculation of the false-alarm probability, we tested the reliability of the \ion{O}{i} lag recovery for the specific sampling and noise properties of the campaign by performing a lag-injection test. We created a simulated \ion{O}{i}$\,\lambda8446$ light curve as a delayed and noisy copy of the \Lya{} light curve, adopting an input lag of 2.5 days, the observed \ion{O}{i} sampling, and the measured \ion{O}{i} uncertainties. Applying the same FR and RSS analysis as for the real data, we recovered a median lag of $2.5^{+0.3}_{-0.5}$ days. This demonstrates that the HST campaign length and daily cadence are sufficient to reliably recover the lags quoted in Sect.~\ref{sec:res_response_OI8446}.
\end{document}